\documentclass{iaus}
\usepackage{graphicx}
\usepackage{amsmath}
\usepackage{amssymb}

\title[IAU Symposium 286.~~Comparative magnetic minima] 
{Stellar activity cycles in a model for magnetic 
flux generation and transport}

\author[Emre I\c{s}\i k]   
{Emre I\c{s}\i k}

\affiliation{Department of Physics, Faculty of Science \& Letters, 
Istanbul K\"ult\"ur University \\
34156, Bak\i rk\"oy, Istanbul, Turkey \\ email: {\tt e.isik@iku.edu.tr} }

\pubyear{2011}
\volume{286}  
\jname{Comparative magnetic minima: characterizing quiet times in the Sun 
and stars}
\editors{A.C. Editor, B.D. Editor \& C.E. Editor, eds.}
\begin{document}

\maketitle

\begin{abstract}
We present results from a model for magnetic flux generation and transport 
in cool stars and a qualitative comparison of models with observations. 
The method combines an $\alpha\Omega$-type dynamo at the 
base of the convection zone, buoyant rise of magnetic flux tubes, and a 
surface flux transport model. Based on a reference model for the Sun, 
numerical simulations were carried out for model convection zones of 
G- and K-type main sequence and subgiant stars. We investigate magnetic cycle 
properties for stars with different rotation periods, convection zone depths, 
and dynamo strengths. For a Sun-like star with $P_{\rm rot}$=9~d, we find that 
a cyclic dynamo can underly an apparently non-cyclic, 'flat' surface activity, as 
observed in some stars. 
For a subgiant K1 star with $P_{\rm rot}$=2.8~d the long-term activity 
variations resemble the multi-periodic cycles observed in V711 Tau, 
owing to high-latitude flux emergence, 
weak transport effects and stochastic processes of flux emergence. 

\end{abstract}

\firstsection 

\section{Introduction}


Magnetically active stars are important to better understand the limits and 
the behaviour of the solar dynamo, in addition to how it used to operate during 
the early stages of solar evolution. The level of activity increases with the 
rotation rate and the fractional depth of the convection zone, up to a 
saturation level for ultra-fast rotators and fully convective stars.
The distribution and coverage of stellar magnetic regions differ significantly 
from the solar patterns as the Rossby number (ratio of the rotation period to 
convective turn-over time) decreases  
(\cite[Strassmeier 2009]{Strass09}). 
Temporal properties of magnetic activity are also of interest when comparing 
the solar cycle with stellar cycles. 
Multi-periodic stellar cycles have also been 
reported (\cite[Ol\'ah et al. 2009]{olah09}). The present paper presents 
a model for the generation and transport of stellar magnetic fields 
(\cite[I\c{s}\i k, Schmitt \& Sch\"ussler 2007, 2011]{isik07, isik11}) and 
discusses possible magnetic flux transport mechanisms in active stars, 
in comparison with observations of stellar activity cycles. 

\section{The model}

We have recently developed a three-part model for solar and stellar magnetic flux 
generation and transport 
(\cite[I\c{s}\i k, Schmitt \& Sch\"ussler 2007, 2011]{isik07, isik11}). 
In the first part, a one-dimensional mean-field $\alpha\Omega$ dynamo 
with a saturation mechanism related to buoyant flux removal is set up at the 
position of the overshoot region at the base of the convection zone 
(\cite[Schmitt \& Sch\"ussler 1989]{ss89}). 
For faster rotators, the strength of the $\alpha$-effect is assumed to scale 
with the rotation rate. This is identical to the assumption that the 
differential rotation rate scales with the activity level and thus with the rotation rate, 
as suggested recently by \cite{saar2011}, based on observations of a 
homogeneous sample of cool stars. In the second part, 
time-latitude distribution of the dynamo-generated toroidal field determines 
the probability distribution that a flux tube would become 
unstable at a given latitude and time. The tubes are assumed to have properties 
given by the criteria 
for magnetic buoyancy instability at a given latitude. The emergence 
latitudes and tilt angles of BMRs are determined by numerical simulations 
of flux tube rise through the convection zone. These results are used 
as input for the flux transport model at the surface (the third part), 
including effects of differential rotation, meridional flow, horizontal and radial 
diffusion. The number of emerging BMRs (per activity cycle) is scaled up with the 
rotation rate and their areas are determined randomly, following a 
probability distribution set by the Sun-like area distribution, 
$N(A)\propto A^{-2}$ . 
The details of the model construction are described by \cite{isik11}. 

\section{Sun-like stars: disappearance of cycles}

For the Sun-like models, the cycle period decreases with the rotation rate, 
owing to stronger $\alpha$-effect. For $P_{\rm rot}\gtrsim 10$~d, the polar 
fields become stronger with increasing rotation rate, 
owing to increasing frequency of flux emergence and tilt angles, 
until another effect starts to set the stage: 
the increasing dynamo strength (the $\alpha$-effect), thus the increasing 
cycle frequency. 
For $P_{\rm rot}\lesssim 10$~d polar regions undergo field reversals 
with increasingly high frequency, so that the peak polar fields do not reach the 
field strengths expected from the emerging magnetic flux. 
For most of these cases the magnetic flux variations show clear cycles. 

\begin{figure}
\begin{center}
\includegraphics[width=.9\linewidth]{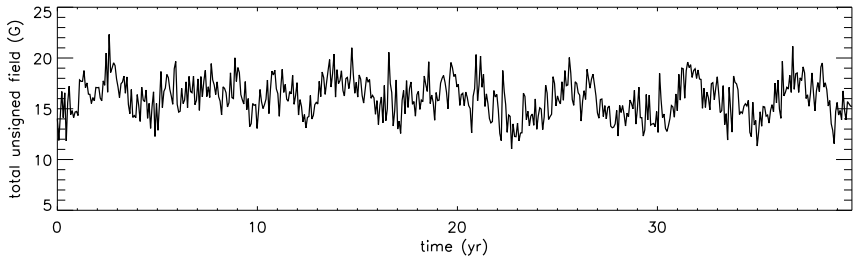}
\caption{Variation of total unsigned magnetic flux for the Sun-like model 
with $P_{\rm rot}$=9~d. The right axis shows the corresponding field strength.}
\label{fig:9d}
\end{center}
\end{figure}

For $P_{\rm rot}=9$~d, the level of total magnetic flux 
is considerably higher than the solar values, but the underlying periodic 
dynamo cycle is 'obscured' by the combined effects of rise and surface transport of 
magnetic flux (Fig.~\ref{fig:9d}). The reason for the 'invisible' stellar cycle is that 
during 'magnetic minima' the magnetic flux at high latitudes become comparable 
to that of low latitudes during activity maxima. 
This intermediate rotator case represents a moderately active, but non-cyclic 
(or weakly cyclic) configuration (in fact, there is a weak cycle with 
$\sim 6$-yr period). The model indicates a possibility that such 
stars may not be in a Maunder minimum state and are still observed as non-cyclic. 
Observational evidence on the existence of such stars already supports 
this possibility (\cite{HallLockwood04}). 

\section{K stars: from regular to fluctuating cycles}

\cite{isik11} have applied the Sun-like model to K-type stars with 
different radii. The first case is a K0-type main-sequence star, rotating 13 times 
faster than the Sun. Similar to the Sun-like star with the same rotation 
rate, large tilt angles lead to formation of activity belts, but in this case 
as shifted towards the poles, because of an additional geometric effect. 
Buoyant flux tubes experience a substantial Coriolis acceleration in the rotating 
frame and rise almost parallel to the rotation axis in fast rotators. 
The larger the fractional depth of the convection zone, the higher the emergence 
latitudes, because the latitude difference between the initial and final latitudes 
of the rising flux tube is larger. 

The second model simulates a K1IV-type fast rotator, with the mass, 
radius, and the rotational period of the active subgiant component of the 
close binary system V711 Tau (=HR 1099). The deep-seated dynamo pattern 
and the surface emergence are highly separated in latitude, owing to rapid 
rotation and convection zone geometry, as explained above. 
\begin{figure}
\begin{center}
 \includegraphics[width=.9\linewidth]{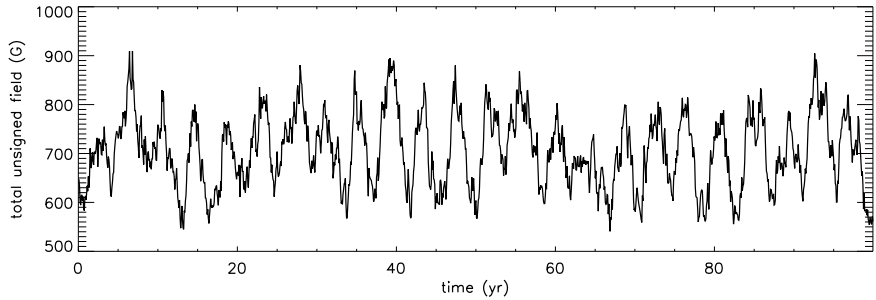} 
 \caption{Variation of the surface-integrated unsigned magnetic field 
   strength for the K1IV-type fast rotator with $P_{\rm rot}$=2.8 d.}
   \label{fig:K1flux}
\end{center}
\end{figure}
The variation of the total unsigned magnetic field for 100 years is shown in 
Fig.~\ref{fig:K1flux}. The random component in the model, 
ie., quasi-random latitudes and areas of BMRs at emergence, become significant 
for the polar regions, where the timescales for differential rotation, 
meridional flow, and the turbulent magnetic diffusion are too long to have 
an effect on the rapidly changing magnetic flux pattern. 
\begin{figure}
\begin{center}
 \includegraphics[width=.45\linewidth]{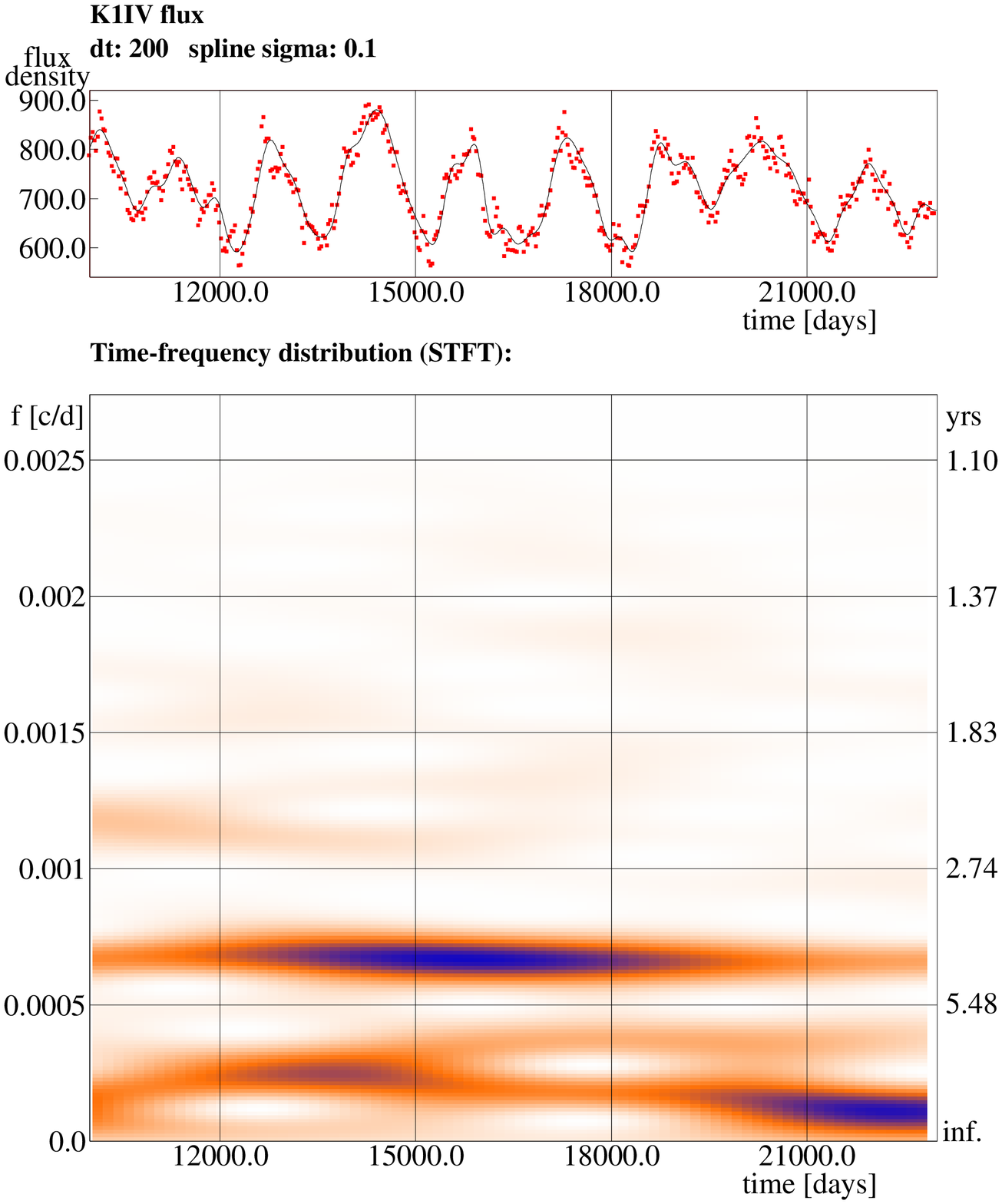} 
 \includegraphics[width=.45\linewidth]{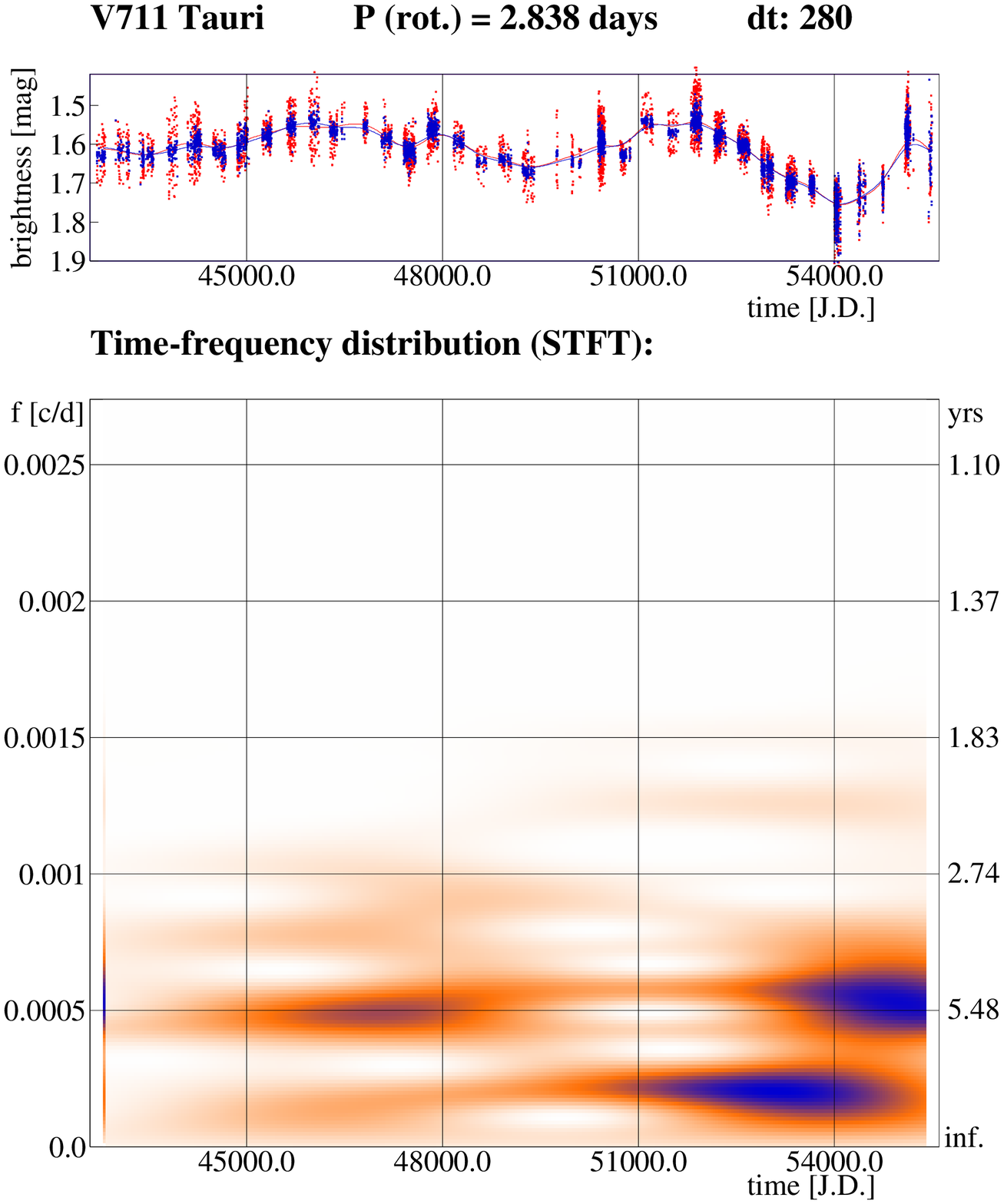} 
 \caption{Time-frequency maps of the theoretical model of K1IV star 
 with $P_{\rm rot}=2.8$~d (\emph{left panel}, based on total unsigned 
 magnetic flux) and V711 Tau (\emph{right panel}, based on stellar brightness). 
 For V711 Tau, the data are from \cite{olah09} extended to 2011 by new measurements from the Vienna APT (\cite[Strassmeier et al. 1997]{strass97}).}
   \label{fig:STFT}
\end{center}
\end{figure}
The ineffective surface transport and the highly confined, 
highly frequent BMR emergence thus lead to cycle-to-cycle (and long-term) 
fluctuations in total magnetic flux. Preliminary time-frequency 
analysis (\cite[Ol\'ah 2011]{olah11}) indicates an interesting correspondence 
between the models and the long-term photometric observations of V711 Tau, 
as shown in Fig.~\ref{fig:STFT}. 
The dynamo cycle period in the model is about 4 years and the short-term 
cycle of V711 Tau is about 5 years. A longer-term 'cycle' shifts its period 
between 10-20 years in the model, whereas for V711 Tau a similar shifting 
cycle is observed between 9-18 years. 
A longer-term simulation for the K1IV model (Fig.~\ref{fig:K1flux})
shows similar periods. However, it should be noted that (a) the average 
brightness on the one hand is compared with the total magnetic flux on the 
other, and (b) the relative amplitudes of the short- and long-term cycles 
are not similar in both cases. 

\section{Discussion}

It is clear that the relationships between stellar dynamo mechanisms, emergence 
patterns, and surface transport processes are non-trivial. The effect of rapid 
rotation on rising magnetic flux tubes in stellar convection zones can lead to  
large departures of observable patterns from the deep-seated fields. 
Surface flux transport processes cause additional complications and indicate 
various possibilities when interpreting the observed activity patterns. Apart from 
several unknowns about active cool stars, we conjecture that three parameters are 
likely to have important effects in shaping stellar cycles: 
the cycle frequency, the range of latitudes of BMR emergence, and the 
tilt angles. 

Multiple cycles observed in active stars 
(\cite[Ol\'ah et al. 2009]{olah09}), such as V711 Tau in 
Fig.~\ref{fig:STFT} (right panel), do not necessarily owe their existence to 
periodic events taking place in the stellar interior. In the theoretical model, the 
short-term cycle ($\sim4$ yrs) is simply determined by the mono-periodic 
dynamo in the interior. 
The long-term cycle signal in Fig.~\ref{fig:STFT} (left panel) is caused by the 
stochastic nature of flux emergence. 
Although there are several degrees of freedom in modelling magnetic fields in 
stellar interiors, parallel investigation of nonlinear flux-transport 
dynamo models and observational results will certainly be useful when 
estimating the relevant physical mechanisms. 

\section*{Acknowledgement}

The author is grateful to Katalin Ol\'ah for the data, time-frequency analysis and 
useful discussions.

\end{document}